\RequirePackage{fix-cm}
\documentclass[twocolumn,epjc3]{svjour3}     

\smartqed  
\usepackage{graphicx}
\usepackage{amsmath}
\usepackage{cuted}
\DeclareMathOperator{\arcsinh}{arsinh}

\begin{document}

\title{Semiclassical two-step model for ionization of the hydrogen molecule by strong laser field
}


\author{N. I. Shvetsov-Shilovski$^{1,a}$, M. Lein$^{1}$, and K. T\H{o}k\'esi$^{2,3}$}

\institute{Institut f\"{u}r Theoretische Physik, Leibniz Universit\"{a}t Hannover, D-30167, Hannover, Germany
					\and
					Institute for Nuclear Research, Hungarian Academy of Sciences, H-4001 Debrecen, Hungary
					\and
					ELI-HU Nonprofit Ltd., Dugonics t\'er 13, 6720 Szeged, Hungary
}
              
\thankstext{e1}{e-mail: nikolay.shvetsov@itp.uni-hannover.de}

\date{\today}

\maketitle

\begin{abstract}
We extend the semiclassical two-step model for strong-field ionization that describes quantum interference and accounts for the Coulomb potential beyond the semiclassical perturbation theory to the hydrogen molecule. In the simplest case of the molecule oriented along the polarization direction of a linearly polarized laser field, we predict significant deviations of the two-dimensional photoelectron momentum distributions and the energy spectra from the case of atomic hydrogen. Specifically, for the hydrogen molecule the electron energy spectrum falls off slower with increasing energy, and the holographic interference fringes are more pronounced than for the hydrogen atom at the same parameters of the laser pulse.  
\end{abstract}

\section{Introduction}
Strong-field physics focuses on the interaction of intense laser radiation with atoms and molecules. This interaction leads to such phenomena as above-threshold ionization (ATI), formation of the high-energy plateau in the electron energy spectrum, generation of high-order harmonics, sequential and nonsequential double and multiple ionization, etc. (see Refs.~\cite{BeckerRev2002,MilosevicRev2003,FaisalRev2005,FariaRev2011} for reviews). Among the main theoretical approaches used in strong-field laser-atom physics are the direct numerical solution of the time-dependent Schr\"odinger equation (TDSE) (see, e.g., Refs.~\cite{Muller1999,Bauer2006,Madsen2007}), the strong-field approximation (SFA) \cite{Keldysh1964,Faisal1973,Reiss1980}, and the semiclassical models (see, e.g., Refs.~\cite{Linden1988,Gallagher1988,Corkum1989,Corkum1993}) applying a classical description of an electron after it has been released from an atom, e.g., by tunneling ionization \cite{Dau3,PPT,ADK}. 

However, ionization of a molecule by a strong laser pulse is much more complicated than the same process for an atom. This is due to the presence of extra degrees of freedom (nuclear motion), the corresponding time scales, and complicated shape of the electronic orbitals. In order to fully consider all the features of the molecular ionization, the TDSE in three spatial dimensions should be solved. This is a very difficult task, which is possible only for the simplest molecules and under selection of the most releavant degrees of freedom (see Refs.~\cite{Palacios2006,Saenz2014}). Therefore, approximate semianalytic models that can provide the basic physical insight are very important for theoretical description of molecules in strong laser fields. The molecular strong-field approximation (MO-SFA) \cite{Faisal2000,Madsen2004} and the molecular tunneling theory, i.e., the molecular Ammosov-Delone-Krainov (MO-ADK) theory \cite{Tong2002}, are the most widely known examples of such models. MO-SFA is a generalization of the atomic SFA onto the case of molecules. In turn, the MO-ADK is the extension of the Perelomov-Popov-Terent'ev \cite{PPT} or ADK \cite{ADK} tunneling formulas. In both MO-SFA and MO-ADK (as well as in the SFA and ADK) ionization is described as a transition from an initial field-free state to a Volkov state (the wave function for an electron in an electromagnetic field). As the result, intermediate bound states and the Coulomb interaction in the final state are completely neglected in the MO-SFA and MO-ADK. 

In contrast to this, the semiclassical approaches allow to account for the effects of the Coulomb potential. In addition to this, the trajectory-based models do not usually require as high computational costs as the numerical solution of the TDSE. Finally, by analyzing classical trajectories these models are able to illustrate the mechanism underlying the strong-field phenomena of interest. For these reasons, the semiclassical models accounting for the ionic potential has become one of the powerful tools of the strong-field physics. However, presently there are only a few studies applying these models to describe strong-field ionization of molecules, see, e.g., Refs.~\cite{Graefe2013,Liu2016,Liu2017}. 

Until recently, trajectory-based models have not been able to account for quantum interference and to describe interference effects in photoelectron spectra and momentum distributions. This drawback has been overcome by a quantum trajectory Monte Carlo (QTMC) \cite{Li2014} and semiclassical two-step (SCTS) \cite{Shvetsov2016} models. In these models each classical trajectory is associated with a certain phase, and the contributions of all the trajectories leading to a given final (asymptotic) momentum are added coherently. In the SCTS model the phase is calculated from the semiclassical expression for the matrix element of the quantum mechanical propagator. Therefore, the SCTS model accounts for the Coulomb potential beyond the semiclassical perturbation theory. The phase used in the QTMC model can be viewed as an approximation to the full semiclassical phase of the SCTS. As the result, the SCTS model yields better agreement with the direct numerical solution of the TDSE than previously developed QTMC model, see Ref.~\cite{Shvetsov2016}. While the QTMC model was applied to strong-field ionization of molecules (see Refs.~\cite{Liu2016,Liu2017}), the SCTS model has not been generalized to the molecular case so far.

In this paper we extend the SCTS model to molecular hydrogen. We calculate two-dimensional momentum distributions, energy spectra, and angular distributions of ionized electrons, and compare them to the case of atomic hydrogen. We reveal significant differences in momentum distributions and electron energy spectra. Then we compare our molecular SCTS to the predictions of the QTMC model for molecules.

The paper is organized as follows. In Sec.~II we formulate the SCTS model for the hydrogen molecule. In Sec.~III we compare the results of the SCTS model for the H$_{2}$ molecule and the H atom. The conclusions are given in Sec.~IV. Atomic units are used throughout the paper unless indicated otherwise.   
                 
\section{Semiclassical two-step model for H$_2$ molecule}

Any semiclassical approach to a strong-field process is based on propagation of an ensemble of classical trajectories that obey Newton's equation of motion. We treat the electric field of the laser pulse $\vec{F}\left(t\right)$ and the ionic potential $V\left(\vec{r},t\right)$ on equal footing, and, therefore, the equation of motion reads as:
\begin{equation}
\frac{d^{2}\vec{r}}{dt^2}=-\vec{F}\left(t\right)-\vec{\nabla}V\left(\vec{r},t\right).
\label{newton}
\end{equation}
For the H$_2$ molecule the ionic potential reads as
\begin{equation}
V\left(\vec{r}\right)=-\frac{Z_{1}}{\left|\vec{r}-\vec{R}/2\right|}-\frac{Z_{2}}{\left|\vec{r}+\vec{R}/2\right|},
\label{molpot}
\end{equation}
where $\vec{R}$ is the vector that points from one nucleus to another. Here we assume that the origin of the coordinate system is placed in the center of the molecule. Following Ref.~\cite{Liu2017}, we choose the effective charges $Z_{1}$ and $Z_{2}$ to be equal to 0.5 a.u. 

In order to find the  trajectory $\vec{r}\left(t\right)$ and momentum $\vec{p}\left(t\right)$ by integrating Eq.~(\ref{newton}), we need to specify the initial velocity of the electron and the starting point of its trajectory. As in many trajectory-based models, we assume that the electron starts with zero initial velocity along the laser field $v_{0,z}=0$, but can have a nonzero initial velocity $\vec{v}_{0,\perp}=v_{0,x}\vec{e}_x+v_{0,y}\vec{e}_y$ in the perpendicular direction (We imply that the field is polarized along the z axis). Various approaches are used to obtain the starting point of the trajectory, i.e., the tunnel exit point $z_{e}$. For the H atom the tunnel exit can be found using the separation of the tunneling problem for the Coulomb potential in parabolic coordinates \cite{Dau3}. However, in the case of the H atom we use the simplest formula for $z_{e}$ neglecting the Coulomb potential, i.e., assuming triangular potential barrier formed by the laser field and the ground state energy:
\begin{equation}
r_{e}\left(t_0\right)=\frac{I_p}{F\left(t_0\right)},
\label{triang}
\end{equation}
where $I_p$ is the ionization potential, $r_{e}\left(t_0\right)=\left|z_{e}\left(t_0\right)\right|$, and the sign of $z_{e}\left(t_0\right)$ is chosen the ensure that the electron tunnels out from under the barrier in the direction opposite to the laser field $\vec{F}\left(t_0\right)$ at the time of ionization. For the H$_{2}$ molecule we use and compare two different approaches to determine $z_{e}$. The first one applies Eq.~(\ref{triang}) neglecting the molecular potential, and the second approach considers the potential barrier formed by the molecular potential and the electric field of the laser \textit{in a 1D cut along the field direction}. This latter approach is often called the field direction model (FDM), see \cite{Shvetsov2012}. Thus in the FDM the tunnel exit point is found from the equation:
\begin{equation}
V\left(\vec{r}\right)+F\left(t_0\right)z_{e}=-I_{p}. 
\label{fdm}
\end{equation}  
The electron trajectory $\vec{r}\left(t\right)$ is completely determined by the ionization time $t_0$ and initial transverse velocity $\vec{v}_{0,\perp}$. However, in contrast to the atomic case, for ionization of a molecule it is not obvious how to distribute the times of ionization and the initial transverse momenta.  

The two main approaches to this problem are presently used in semiclassical simulations of the ATI in molecules: Molecular quantum-trajectory Monte-Carlo model (MO-QTMC) \cite{Liu2016,Liu2017} and partial Fourier transform approach for molecules (MO-PFT), see Refs.~\cite{Ivanov2011,Liu20162,Liu2017}. The MO-QTMC approach is based on the MO-SFA, whereas the MO-PFT introduces the wave function in mixed representation and applies the Wentzel-Kramers-Brillouin (WKB) approximation. For the parameters of interest both approaches yield similar results when combined with the QTMC model (see Ref.~\cite{Liu2017}). In the present paper we use the MO-PFT model. This model is based on the partial Fourier transform (PFT) approach for atoms \cite{Ivanov2010}. In order to make the presentation self-contained, here we repeat the main points of the PFT and MO-PFT. 

The mixed representation wave function $\Pi\left(p_x, p_y, z\right)$ of an electron in a zero-range potential and a static electric field that points in the direction of the z axis is given by (see Ref.~\cite{Ivanov2010}):
\begin{align} 
& \Pi\left(p_x, p_y, z\right)=\Pi\left(p_x, p_y, z_{0}\right)\sqrt{\frac{p_z\left(z_0\right)}{p_z\left(z\right)}}\nonumber\\
& \times\exp\left\{i\left[S\left(p_x,p_y,z\right)-S\left(p_x,p_y,z_{0}\right)\right]\right\}.
\label{atom1}
\end{align}
Here $p_z\left(z\right)=\left|\partial S\left(p_x,p_y,p_z\right)/ \partial z\right|$ is the electron momentum in the z direction, $S\left(p_x,p_y,z\right)$ is the classical action, and $z_0<0$ is the point (for $F>0$) in the classically forbidden region, where the one-dimensional WKB solution for the motion along the z axis is matched to the wave function of the initial bound state. We note that since $z_0$ is not a turning point, the wave function and its derivatives are continuous at $z_0$. The exponential of Eq.~(\ref{atom1}) can be found by integration of the Hamilton-Jacobi equation:
\begin{align}
& S\left(p_x,p_y,z\right)-S\left(p_x,p_y,z_{0}\right)\nonumber\\
& =\frac{1}{3F}\left(2E'-2Fz\right)^{3/2}-\frac{1}{3F}\left(2E'-2Fz_{0}\right)^{3/2}.
\label{action2}
\end{align}
Here $E'=-\left(I_{p}+p_{\perp}^2/2\right)$ and $p_{\perp}=\sqrt{p_x^2+p_y^2}$. Assuming that $\left|z_{0}\right|<<\left|z_{e}\right|$, where $z_e=-I_{p}/F<0$ is the exit point from the triangular barrier, we expand the expression in the right-hand side of Eq.~(\ref{action2}) in powers of $z_0$ up to the first order:
\begin{equation}
S\left(p_x,p_y,z_{e}\right)-S\left(p_x,p_y,z_{0}\right)=i\frac{\kappa^3}{3F}+i\frac{\kappa p_{\perp}^2}{2F}-i\kappa\left|z_{0}\right|,
\label{action3}
\end{equation}
where $\kappa=\sqrt{2I_{p}}$. Inserting this expression into Eq.~(\ref{atom1}) and replacing the momentum $p_z\left(z_0\right)$ with $\kappa$, we obtain the following expression for the wave function in mixed representation (see Ref.~\cite{Ivanov2010}): 
\begin{align} 
&\Pi\left(p_x, p_y, z\right)=\Pi\left(p_x, p_y, z_{0}\right)\sqrt{\frac{\kappa}{p_z\left(z\right)}}\nonumber\\
&\times \exp\left[-\frac{\kappa^3}{3F}-\frac{\kappa p_{\perp}^2}{2F}+\kappa\left|z_{0}\right|\right].
\label{atom2}
\end{align}
For the H$_{2}$ molecule the bound-state orbital is the bonding superposition of the two 1s atomic orbitals at the atomic centers (see, e.g., Ref. \cite{McQuarrie}). 
\begin{align}
\Psi_{H_{2}}\left(\vec{r}\right)&=\frac{1}{\sqrt{2\left(1+S_{OI}\right)}}\left[\psi_{atom1}\left(\vec{r}-\vec{R}/2\right)\right.\nonumber \\
&+\left.\psi_{atom2}\left(\vec{r}+\vec{R}/2\right)\right],
\label{homo}
\end{align} 
where $S_{OI}$ is the atomic overlap integral. In what follows we omit the coefficient before the sum of two atomic orbitals in Eq.~(\ref{homo}), since it has no effect on the shape of the momentum distributions that we are interested in. The partial Fourier transform of this molecular orbital reads as (see~Ref.~\cite{Liu20162}):\\
\begin{strip}
\\
\rule{0.485\textwidth}{0.1pt}
\hspace{-0.24cm}
\rule{0.1pt}{0.25cm}
\begingroup\abovedisplayskip=0.75cm \belowdisplayskip=0.75cm
\begin{align}
\Pi_{H_2}\left(p_x,p_y,z\right)&=\exp \left(-\frac{i}{2}p_xR\sin\theta_{m}\cos\varphi_{m}-\frac{i}{2}p_yR\sin\theta_{m}\sin\varphi_{m}\right)\Pi_{atom1}\left(p_x,p_y,z-\frac{R}{2}\cos\theta_{m}\right)\nonumber\\
& +\exp \left(\frac{i}{2}p_xR\sin\theta_{m}\cos\varphi_{m}+\frac{i}{2}p_yR\sin\theta_{m}\sin\varphi_{m}\right)\Pi_{atom2}\left(p_x,p_y,z+\frac{R}{2}\cos\theta_{m}\right),
\label{mopft}
\end{align}
\endgroup
where $\theta_{m}$ and $\varphi_{m}$ are the polar and azimuthal angles of the molecular axis, respectively (see Fig.~1). Inserting the partial Fourier transforms of the 1s orbitals (see Ref.~\cite{Ivanov2010}) into Eq.~(\ref{mopft}) and taking into account that the new matching points $z_1$ and $z_2$ for atoms 1 and 2 are $z_{1,2}=z_{0}\pm R\cos\theta_{m}/2$, we arrive at the following expression for the wave function of the H$_2$ molecule in mixed representation just beyond the tunnel exit:
\begin{align}
\Pi\left(p_{x},p_{y},z_{e}\right)&\sim \left\{\!\begin{aligned}
& \exp\left(-\frac{i}{2}p_xR\sin\theta_{m}\cos\varphi_{m}-\frac{i}{2}p_yR\sin\theta_{m}\sin\varphi_m\right)\exp\left(-\frac{1}{2}\kappa R \cos \theta_{m} \right)\nonumber\\
& +\exp\left(\frac{i}{2}p_xR\sin\theta_{m}\cos\varphi_{m}+\frac{i}{2}p_yR\sin\theta{m}\sin\varphi_{m}\right)\exp\left(\frac{1}{2}\kappa R \cos \theta_{m} \right)
\end{aligned}\right\}\nonumber\\
&\times \exp\left[-\frac{\kappa^3}{3F}-\frac{\kappa\left(p_{x}^2+p_{y}^2\right)}{2F}\right].
\label{distr} 
\end{align}
\end{strip}
\begin{figure}[h]
\begin{center}
\includegraphics[width=0.45\textwidth,trim={0.5cm, 2.0cm, 3.5cm, 0.5cm},clip]{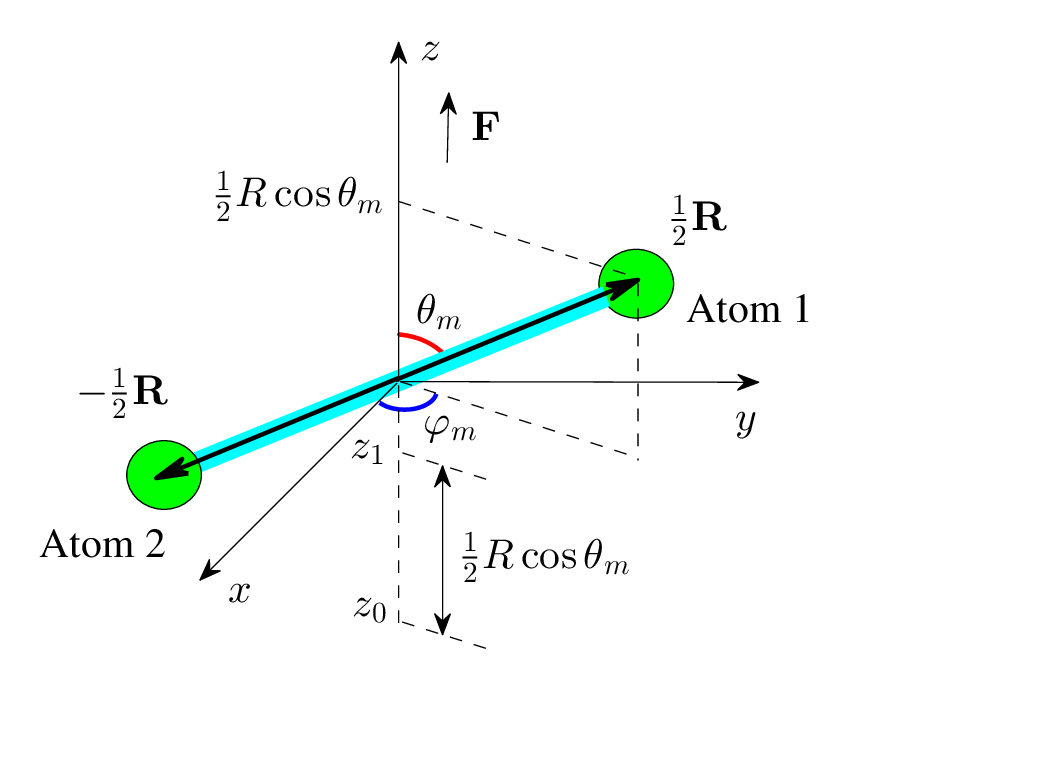}
\end{center}
\caption{The sketch of the hydrogen molecule. The thick blue line depicts the molecular axis. Its orientation is determined by the polar $\theta_{m}$ and azimuthal $\varphi_{m}$ angles. The positions of the first and the second atoms are given by the vectors $-\vec{R}/2$ and $\vec{R}/2$, respectively.  The laser field $\vec{F}$ points towards z direction. The longitudinal displacement $R\cos\theta_{m}/2$ and the new matching point $z_1$ for atom 1 are shown on the z-axis.}
\label{fig:01}       
\end{figure}
Following Ref.~\cite{Liu2017}, we use expression (\ref{distr}) as a complex amplitude for ionization at $t_0$ and $\vec{v}_{0,\perp}$, without adding prefactors. Thus, we put $F=F\left(t_0\right)$, $p_{x}=v_{0,x}$, and $p_{y}=v_{0,y}$ in Eq.~(\ref{distr}). Furthermore, we concentrate on the case when the H$_2$ molecule is oriented along the polarization direction of the laser field, i.e., $\theta_{m}=\varphi_{m}=0$. Then the factor in the curly brackets in Eq.~(\ref{distr}) is a constant for a fixed $R$. Therefore, in our simulations we use only the exponential part of the amplitude (\ref{distr}). We use the ionization potential $I_p=0.60$ a.u., what corresponds to the vertical ionization potential of H$_{2}$ molecule, i.e., to the case when the H$_{2}^{+}$ ion has the same internuclear distance $R$ as the neutral hydrogen molecule. Since in our model the positions of the nuclei are fixed, we disregard the vibrational energy levels when calculating the ionization potential. 

For not too short laser pulses the ionized electron is far away from both nuclei at the time instant $t_f$ when the pulse terminates: $r\left(t_f\right)\gg R/2$. Therefore, we can assume that after the end of the pulse the electron moves in the Coulomb field only. Indeed, the potential of Eq.~(\ref{molpot}) tends to $Z/r$ for $r\to\infty$, where $Z=Z_{1}+Z_{2}$. Therefore, the asymptotic momentum $\vec{k}$ of the electron is uniquely determined by its momentum and position at the end of the pulse \cite{Shvetsov2009,Shvetsov2012}. The magnitude of the final momentum can be found from energy conservation: 
\begin{equation}
\frac{k^2}{2}=\frac{p^2(t_f)}{2}-\frac{Z}{r(t_f)},
\label{conserv}
\end{equation}
whereas its orientation is determined by the following expression:
\begin{equation}
\label{mominf}
\vec{k}=k\frac{k\left(\vec{L}\times\vec{a}\right)-\vec{a}}{1+k^2L^2}.
\end{equation}
Here $\vec L=\vec r(t_f)\times\vec p(t_f)$ and $\vec a=\vec p(t_f)\times\vec L - Z\vec r(t_f)/r(t_f)$ are the conserving angular momentum and the Runge-Lenz vector, respectively. For a single-cycle or two-cycle pulse the condition $r\left(t_f\right)\gg R/2$ is not true for a substantial number of trajectories. In this case the asymptotic momenta are to be found by numerical integration of Eq.~(\ref{newton}) up to times substantially exceeding the duration of the pulse.

In order to accomplish the generalization of the SCTS model to the hydrogen molecule, we need to derive the corresponding expression for the phase associated with a classical trajectory. In the case of an arbitrary effective potential $V\left(\vec{r},t\right)$ the SCTS phase reads as:
\begin{align}
& \Phi^{SCTS}\left(t_{0},\vec v_0\right)= - \vec v_0\cdot\vec r(t_0) + I_{p}t_{0} \nonumber\\
& -\int_{t_0}^\infty dt\ \left\lbrace\frac{p^2(t)}{2}+V[\vec{r}(t)]-\vec r(t)\cdot\vec\nabla V[\vec{r}(t)]\right\rbrace\,
\label{Phi_sim}
\end{align}
see Ref.~\cite{Shvetsov2016}. For the molecular potential $V\left(\vec{r}\right)$ [Eq.~(\ref{molpot})] the phase $\Phi\left(t_{0},\vec v_0\right)$ is given by
\begin{align}
& \Phi_{H_{2}}^{SCTS}\left(t_{0},\vec v_0\right)= - \vec v_0\cdot\vec r(t_0) + I_{p}t_{0} \nonumber\\
& -\int_{t_0}^\infty dt\ \left\lbrace\frac{p^2(t)}{2}-\frac{Z_{1}\left(\vec{r}-\vec{R}/2\right)\cdot\left(2\vec{r}-\vec{R}/2\right)}{\left|\vec{r}-\vec{R}/2\right|^3}\right. \nonumber\\
&\left.+\frac{Z_{2}\left(\vec{r}+\vec{R}/2\right)\cdot\left(2\vec{r}+\vec{R}/2\right)}{\left|\vec{r}+\vec{R}/2\right|^3}\right\rbrace\ .
\label{Phi_mol}
\end{align}
It is easy to see that for $r\gg R/2$ 
\begin{align}
&\Phi_{H}^{SCTS}\left(t_{0},\vec v_0\right)\approx - \vec v_0\cdot\vec r(t_0) + I_{p}t_{0} \nonumber\\
& -\int_{t_0}^\infty dt \left\{\frac{p^2\left(t\right)}{2}-\frac{2Z}{r\left(t\right)}\right\}.
\label{SCTSatom}
\end{align}
Equation (\ref{SCTSatom}) coincides with the SCTS phase for the Coulomb potential, $-Z/r$ (see Ref.~\cite{Shvetsov2016}). Assuming that $r\left(t_f\right)\gg R/2 $ we can use the expression for the phase accumulated in the asymptotic interval $\left[t_{f},\infty\right]$ for the Coulomb potential (see Ref.~\cite{Shvetsov2016}):
\begin{equation}
\tilde{\Phi}^{C}_{f}\left(t_{f}\right)=-Z\sqrt{b}\left[ \ln g+\arcsinh\left\{\frac{\vec{r}(t_f)\cdot\vec{p}(t_f)}{g\sqrt{b}}\right\}\right], 
\label{phase_asym}
\end{equation}
where $b=1/2E$ and $g=\sqrt{1+2EL^2}$. Here, in turn, $E$ is the conserving energy of the electron moving along the outgoing Kepler hyperbola. Equation (\ref{phase_asym}) allows to decompose the integral in Eq.~(\ref{Phi_mol}) into two integrals over the intervals $\left[t_0,t_f \right]$ and $\left[t_f,\infty\right]$. The first integral over $\left[t_0,t_f \right]$ can be readily calculated numerically knowing the position vector $\vec{r}\left(t\right)$ and momentum $\vec{p}\left(t\right)$ of the electron. 

In contrast to this, the phase associated with each trajectory in the molecular QTMC model reads as:
\begin{align}
&\Phi_{H_{2}}^{QTMC}\left(t_{0},\vec v_0\right)= - \vec v_0\cdot\vec r(t_0) + I_{p}t_{0} \nonumber\\
& -\int_{t_0}^\infty dt \left\{\frac{p^2\left(t\right)}{2}-\frac{Z_{1}}{\left|\vec{r}-\vec{R}/2\right|}-\frac{Z_{2}}{\left|\vec{r}+\vec{R}/2\right|}\right\},
\label{QTMC}
\end{align} 
see Ref.~\cite{Liu2017}. At large distances Eq.~(\ref{QTMC}) recovers the QTMC phase for the H atom (see Ref.~\cite{Li2014}):
\begin{align}
&\Phi_{H}^{QTMC}\left(t_{0},\vec v_0\right)\approx - \vec v_0\cdot\vec r(t_0) + I_{p}t_{0} \nonumber\\
& -\int_{t_0}^\infty dt \left\{\frac{p^2\left(t\right)}{2}-\frac{Z}{r\left(t\right)}\right\}.
\label{QTMCatom}
\end{align}

For the simulations we apply an importance sampling implementation of both QTMC and SCTS models. Within the importance sampling approach the initial conditions $t^{0}_{j}$ and $v_{0}^{j}$ $\left(j=1...n_{p}\right)$ for all $n_p$ trajectories of the ensemble are distributed according to the square root of the probability [Eq.~(\ref{distr})]. We propagate the trajectories of the ensemble using Eq.~(\ref{newton}) and bin them in cells in momentum space in accord with their asymptotic momenta. The amplitudes that correspond to the trajectories reaching the same cell are added coherently, and the ionization probability reads as:
\begin{equation}
\frac{dR}{d^{3}k}=\left|\sum_{j=1}^{n_p}\exp\left[i\Phi\left(t_{0}^{j},\vec v_0^{j}\right)\right]\right|^2,
\label{prob}
\end{equation}
see Ref.~\cite{Shvetsov2016}. It should be stressed that it is necessary to achieve convergence both with respect to the size of the momentum cell in and the number of the trajectories in the ensemble.  

\section{Results and discussion}
We define a few-cycle linearly polarized laser pulse in terms of a vector potential:
\begin{equation}
\vec{A}=\frac{F_0}{\omega}f\left(t\right)\sin\left(\omega t\right)\vec{e}_{z}.
\label{vecpot}
\end{equation} 
Here $F_0$ is the field strength, $\omega$ is the carrier angular frequency, $\vec{e}_{z}$ is the unit vector in the polarization direction, and $f\left(t\right)$ is the envelope function. The pulse is present between for $0\leq t \leq t_{f}$. Here $t_{f}=nT$, where, in turn, $n$ is the number of cycles within the pulse, and $T=2\pi / \omega$ is the laser period. The electric field can be obtained from the vector potential (\ref{vecpot}) by $\vec{F}=-d\vec{A}/dt$. In our simulations we consider a pulse with $n=8$ optical cycles and use two different envelope functions:
\begin{equation}
f_{1}\left(t\right)=\sin^2\left(\frac{\omega t}{16}\right),
\label{f1} 
\end{equation}
for a sine squared pulse, and 
\begin{equation}
f_2(t) = 
 \begin{cases}
   t/2T,                 & t<2T\\
   1,                    & 2T \leq t < 6T\\
	 \left(8T-t\right)/2T, & 6T \leq t\leq 8T
 \end{cases}
\label{f2}
\end{equation}
for a trapezoidal pulse.  

We first compare the results of the SCTS model for the H atom and the H$_{2}$ molecule. In Fig.~1 we present the two-dimensional (2D) photoelectron momentum distributions in the $\left(k_z, k_{\perp}\right)$ plane for H [panel (a)] and H$_{2}$ [panels (b) and (c)]. The tunnel exit point for the H atom was calculated from Eq.~(\ref{triang}) [triangular potential barrier]. For the H$_2$ molecule we show the momentum distribution calculated for the exit point given by Eq.~(\ref{triang}) [see Fig.~1~(b)], as well as the distribution obtained for $z_e$ found in accord with the FDM [see Fig.~1~(c)]. It is seen that the distributions of Figs.~1~(a) and 1~(b) are rather similar to each other. It should be stressed that the potential of Eq.~(\ref{molpot}) is fully taken into account in Newton's equations of motion and the expression for the phase when calculating of Fig.~1.~(b). These results show that the effects of molecular structure are not pronounced in the photoelectron momentum distributions if the molecular potential is neglected in finding the starting point of the trajectory. This fact is easy to understand taking into account that for the parameters of Fig.~1 the characteristic exit for the triangular barrier $r_0=I_{p}/F_{0}\gg R/2$. Indeed, for the H molecule at the intensity of $2.0\cdot 10^{14}$~W/cm$^2$ $z_{e,c}\approx 7.94$~a.u. (cf. with $R/2=0.71$~a.u.). We note that the distance between the released electron and the nuclei increases further when the electron moves along the trajectory. Therefore, if the tunnel exit point is calculated from Eq.~(\ref{triang}), the ionized electron is actually moving in the same Coulomb potential as in the case of the atomic hydrogen.
\begin{figure}[h]
\begin{center}
\includegraphics[width=0.45\textwidth]{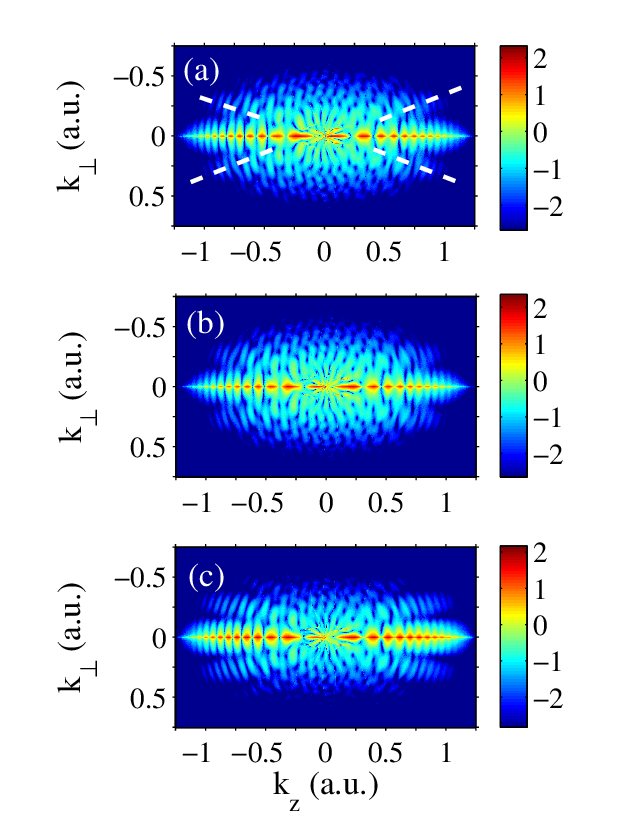}
\end{center}
\caption{The two-dimensional photoelectron momentum distributions for the H atom [panel (a)] and the H$_2$ molecule [panels (b) and (c)] ionized by a laser pulse with a sine square envelope [Eq.~(\ref{f1})], duration of $n=8$ cycles, wavelength of $\lambda=800$ nm, and intensity of $2.0\times10^{14}$ W/cm$^2$. The panels (a) and (b) show the distributions calculated for the exit point given by Eq.~(\ref{triang}). The panel (c) correspond to the exit point found in accord with the FDM [Eq.~(\ref{fdm})]. The holographic fringes are indicated by white dashed lines in panel (a). The distributions are normalized to the total ionization yield. A logarithmic color scale in arbitrary units is used. The laser field is linearly polarized along the $z$ axis.}
\label{fig:1}       
\end{figure}
For the H$_{2}$ molecule and the parameters of Fig.~1 the FDM predicts the tunnel exit point at 5.48~a.u. at the maximum of the laser field. In Fig.~2 we show the tunnel exit points corresponding to the maximum value of the pulse calculated from Eq.~(\ref{triang}) and in accord with the FDM as functions of the laser intensity. It is seen that the exit point from under the triangular potential barrier is larger than the result of the FDM that accounts for the molecular potential [Eq.~(\ref{molpot})]. Moreover, for the parameters of Fig.~1 the FDM with the potential of Eq.~(\ref{molpot}) yields the critical (threshold) field that corresponds to the barrier-suppression regime $F_{C}=0.086$ a.u. This value of the critical field corresponds to the peak intensity of $2.60\cdot10^{14}$ W/cm$^2$ (see Fig~2), which is close to the widely known estimate for the critical field based on the one-dimensional consideration (see Ref.~\cite{BetheBook}):
\begin{equation} 
F_{C}=\frac{\kappa^4}{16Z}
\label{critical}
\end{equation}
that predicts $F_{C}=0.09$ a.u. for a model atom with $I_{p}=0.60$~a.u equal to the one for H$_2$. For these reasons, in what follows we use the FDM for calculation of the exit point for the H$_2$ molecule. 
\begin{figure}
\begin{center}
\includegraphics[width=0.46\textwidth]{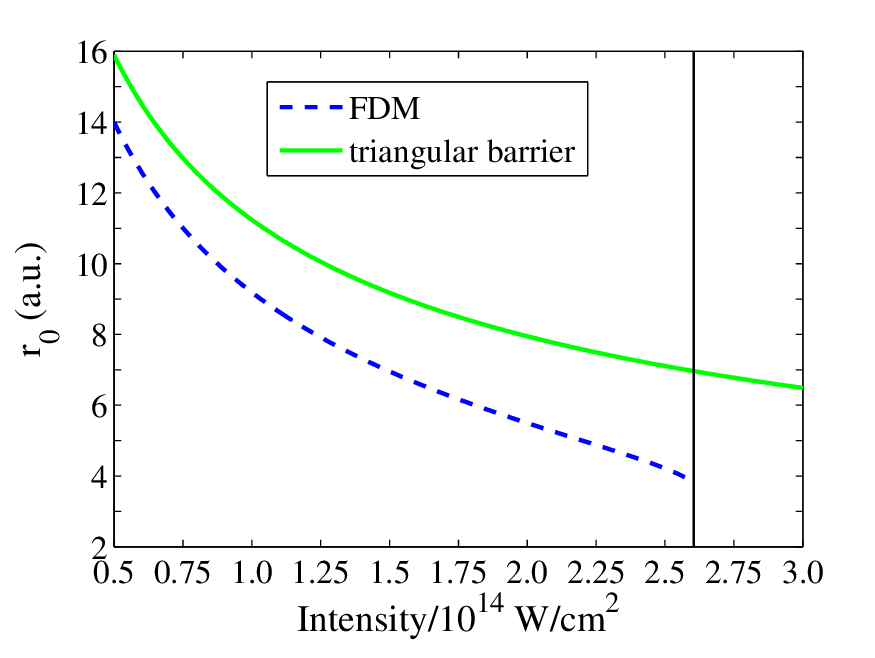}
\end{center}
\caption{The tunnel exit point as a function of laser intensity. The parameters are as in Fig.~1. The blue (dashed) and green (solid) curves correspond to the triangular potential barrier [Eq.~(\ref{triang})] and the FDM model, respectively. The critical (threshold) intensity in the FDM model is shown by a thin black (solid) line.}  
\label{fig:2}       
\end{figure}
The comparison of Figs.~1 (a) and (c) shows that the molecular potential has a pronounced effect on the shape of the momentum distribution and its interference structure. Indeed, the 2D momentum distribution for the H$_2$ molecule is more extended along the polarization direction that the one for the H atom. It is seen that the interference pattern contains well-resolved side lobes (fringes) marked by white (dashed) lines in Fig.~1~(a). These fringes result from holographic interference (see Ref.~\cite{Huismans2011}). It is seen that at the same laser parameters the holographic fringes are more pronounced for the hydrogen molecule than for the H atom. 

The photoelectron energy spectra and angular distributions calculated for H and H$_{2}$ are shown in Figs.~3 (a) and (b), respectively. It is seen that the energy spectrum for the H$_2$ molecule falls off slower than the spectrum for the H atom. This is a direct consequence of the fact that the 2D momentum distribution for the hydrogen molecule is more extended along the polarization direction, where the signal is maximal. Accordingly, the angular distribution for H$_2$ is more aligned along the polarization axis than that for the H atom, see Fig~3~(b). 
\begin{figure}[h]
\begin{center}
\includegraphics[width=0.47\textwidth]{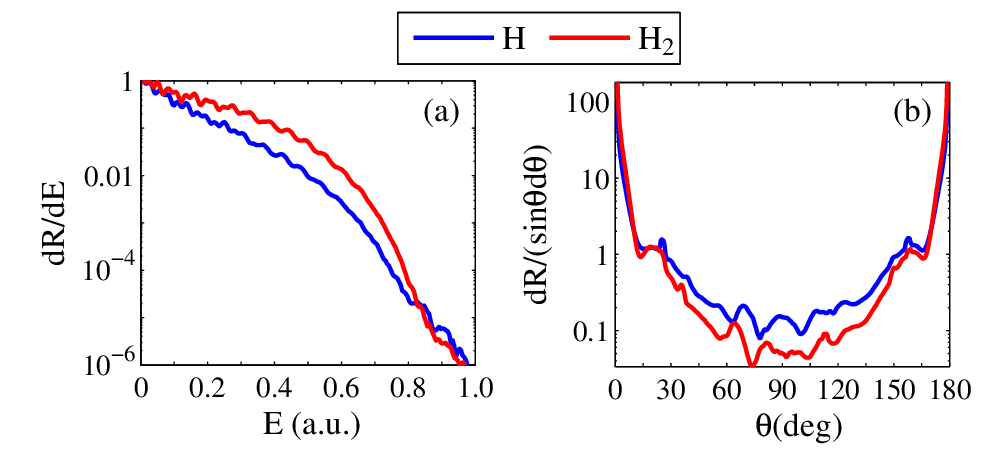}
\end{center}
\caption{Energy spectra (a) and and angular distributions (b) of the photoelectrons for ionization of H and H$_2$. The parameters are the same as in Fig.~1. The energy spectra are normalized to the peak value, and the angular distributions are normalized to the total ionization yield.}
\label{fig:3}       
\end{figure} 
Up to this point we have discussed the results of the SCTS model. It is instructive to compare these results with the predictions of the QTMC model. In Figs.~4 (a)-(d) we present the 2D electron momentum distributions for the H$_2$ molecule calculated in accord with the QTMC [panels (a) and (c)] and the SCTS [panels (b) and (d)]. The first and the second row of Fig.~4, i.e., the panels [(a),(b)] and [(c),(d)] show the results for the sine squared [Eq.~(\ref{f1})] and trapezoidal [Eq.~(\ref{f2})] pulse, respectively. Figures 5 (a)-(d) show the magnification of the distributions of Figs.~4~(a)-(d) for $\left|k_{z}\right|$, $\left|k_{\perp}\right|<0.35$~a.u. It is seen from Figs.~5 (a) and (b) that in their low-energy part the 2D momentum distributions display fan-like interference structures. For the trapezoidal pulse the interference structure in the low-energy part of the distributions consists of a number of blobs on a circle of the radius $k\approx 0.30$~a.u. [see Figs.~5 (c) and (d)]. These structures are similar to that of Ramsauer-Townsend diffraction oscillations were thoroughly studied in Refs.~\cite{Arbo2006,Arbo2008,Moshammer2009,Arbo2010}. It is seen that similar to the atomic case \cite{Shvetsov2016}, the QTMC predicts for the H$_2$ molecule fewer nodal lines in the low-energy interference structure than the SCTS model. As in the case of atomic hydrogen, we attribute this fact to the underestimation of the Coulomb interaction in the QTMC treatment of the interference phase.    
\begin{figure}[h]
\begin{center}
\includegraphics[width=0.48\textwidth]{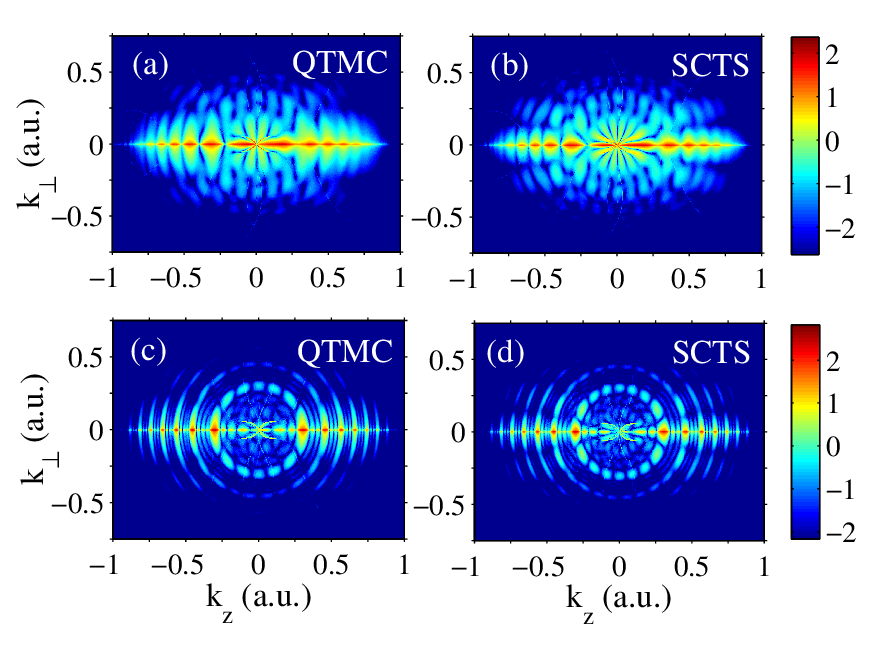}
\end{center}
\caption{The two-dimensional photoelectron momentum distributions for H$_2$ ionized by a laser pulse with an intensity of $1.2\times10^{14}$ W/cm$^2$. The rest of parameters are the same as in Fig.~1. The left column [panels (a) and (c)] show the predictions of the QTMC model. The right column [panels (b) and (d)] displays the results SCTS model. Panels (a,b) and (c,d) correspond to the sine squared [Eq.~(\ref{f1})] and trapezoidal [Eq.~(\ref{f2})] pulse, respectively. The distributions are normalized to the total ionization yield. A logarithmic color scale in arbitrary units is used. The laser field is linearly polarized along the $z$ axis.}  
\label{fig:4}       
\end{figure}
\begin{figure}[h]
\begin{center}
\includegraphics[width=0.48\textwidth]{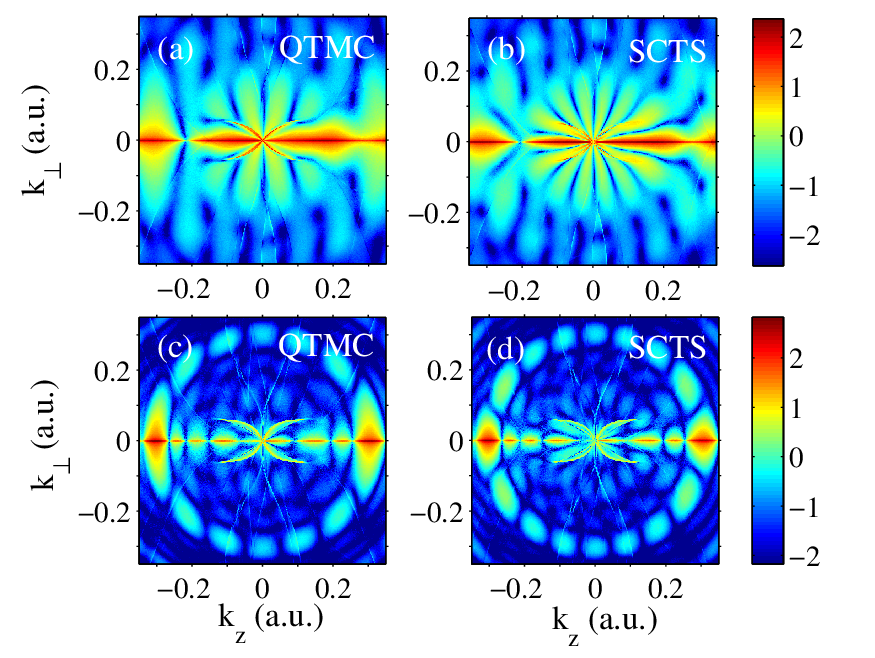}
\end{center}
\caption{Magnification of Fig.~4 for $\left|k_z\right|<0.35$~a.u. and $\left|k_{\perp}\right|<0.3$~a.u.}
\label{fig:5}       
\end{figure}

\section{Conclusions}
In conclusion, we have extended the SCTS model of Ref.~\cite{Shvetsov2016} to ionization of the hydrogen molecule by a strong few-cycle laser pulse. We have restricted ourselves to the simplest case when the H$_2$ molecule is oriented along the direction of a linearly polarized laser field. In our simple implementation of the molecular SCTS model the ionized electron moves in the laser field and in the Coulomb fields of the two fixed atomic nuclei with equal effective charges of 0.5 a.u. 

We have compared the 2D photoelectron momentum distributions calculated for the H$_2$ molecule and the hydrogen atom. We have found that for the same laser parameters the distributions for H$_2$ are more extended along the polarization axis than the ones for H. As the result, the energy spectrum for the H$_2$ molecule falls off slower with electron energy than for the hydrogen atom. Furthermore, the holographic interference fringes in the 2D electron momentum distributions are more pronounced for the hydrogen molecule than for the H atom. 

By comparing with the predictions of the QTMC model for H$_2$ we have found that similar to the atomic case, the QTMC yields fewer nodal lines in the interference structure that is characteristic for the low-energy part of the momentum distributions. As in the case of the H atom, we attribute this to the fact that the QTMC model underestimates the Coulomb interaction in the phase associated with each classical trajectory.    

The present SCTS model for H$_2$ can be straightforwardly extended to an arbitrary orientation of the molecule and polarization of the laser field, as well as to other molecules, including heteronuclear and polyatomic ones. We believe that this further development of the molecular SCTS model will help to understand better the complicated highly nonlinear phenomena originating from the interaction of intense laser radiation with molecules.      
\vspace{12pt}

\noindent This work was supported by the Deutsche Forschungsgemeinschaft (Grant No. SH 1145/1-1) and by the National Research, Development and Innovation Office\\ \mbox{(NKFIH)} Grant KH126886 and partial support by the  ELI-ALPS projects (GOP-1.1.1-12/B-2012-000, GINOP-2.3.6-15-2015-00001).

\section{Author contribution statement}
All authors discussed the results and contributed to the final manuscript. Nikolay Shvetsov-Shilovski performed the calculations.

\end{document}